\documentclass[aps,onecolumn,showpacs,showkeys,nofootinbib]{revtex4}
\usepackage{epsfig}
\usepackage{amsmath}
\usepackage{amsfonts}
\usepackage{amssymb}
\usepackage{graphicx}
\usepackage{colordvi}
\begin{document}

\title{The most general fourth order theory of Gravity at low energy}

\author{A. Stabile\footnote{e -
mail address: arturo.stabile@gmail.com}}

\affiliation{Dipartimento di Ingegneria, Universita'
del Sannio\\Corso Garibaldi, I - 80125 Benevento, Italy}

\begin{abstract}

The Newtonian limit of the most general fourth order gravity is
performed with metric approach in the Jordan frame with no gauge
condition. The most general theory with fourth order differential
equations is obtained by generalizing the $f(R)$ term in the
action with a generic function containing other two curvature
invariants: \emph{Ricci square} ($R_{\alpha\beta}R^{\alpha\beta}$)
and \emph{Riemann square}
($R_{\alpha\beta\gamma\delta}R^{\alpha\beta\gamma\delta}$). The
spherically symmetric solutions of metric tensor yet present
Yukawa-like spatial behavior, but now one has two characteristic
lengths. At Newtonian order any function of curvature invariants
gives us the same outcome like the so-called \emph{Quadratic
Lagrangian} of Gravity. From Gauss - Bonnet invariant one have the
complete interpretation of solutions and the absence of a possible
third characteristic length linked to Riemann square contribution.
From analysis of metric potentials, generated by point-like
source, one has a constraint condition on the derivatives of $f$
with respect to scalar invariants.

\end{abstract}
\pacs{04.25.Nx; 04.50.Kd; 04.40.Nr}
\keywords{Alternative theories of gravity; newtonian and post-newtonian limit; weak field limit.}
\maketitle

\section{Introduction}

In recent years, the effort to give a physical explanation to the
today observed cosmic acceleration \cite{cosmic_acceleration} has
attracted a good amount of interest in $f(R)$-gravity, where $f$
is a generic function of Ricci scalar $R$, considered as a viable
mechanism to explain the cosmic acceleration by extending the
geometric sector of field equations without the introduction of
dark matter and dark energy. Other issues, od astrophysical
nature, as the observed Pioneer anomaly problem \cite{anderson}
can be framed into the same approach \cite{bertolami} and then,
apart the cosmological dynamics, a systematic analysis of such
theories urges at short scale and in the low energy limit.

While it is very natural to extend the theory of General
Relativity (GR) to theories with additional geometric degrees of
freedom, recent attempts focused on the old idea
\cite{old_papers_fR} of modifying the gravitational Lagrangian in
a purely metric framework, leading to higher-order field
equations. Due to the increased complexity of the field equations
in this framework, the main body of works dealt with some formally
equivalent theories, in which a reduction of the order of the
field equations was achieved by considering the metric and the
connection as independent objects \cite{francaviglia}.

In addition, other authors exploited the formal relationship to
scalar-tensor theories to make some statements about the weak
field regime \cite{olmo}, which was already worked out for
scalar-tensor theories \cite{Damour:Esposito-Farese:1992}. Also a
Post-Newtonian parameterization with metric approach in the Jordan
Frame has been considered \cite{clifton}.

In this paper, we show the theory of Gravity induced by a most
general fourth order theory obtained by using all curvature
invariants. Precisely we show for a generic $f(X,Y,Z)$-theory,
where $X\,=\,R$, $Y\,=\,R_{\alpha\beta}R^{\alpha\beta}$ and
$Z\,=\,R_{\alpha\beta\gamma\delta}R^{\alpha\beta\gamma\delta}$
with $R_{\alpha\beta}$ Ricci tensor and
$R_{\alpha\beta\gamma\delta}$ Riemann tensor, the modifications to
standard gravitational mechanics at Newtonian order. From the
usual \emph{small velocity and weak field limit approach}
\cite{PRD_mio} we find the field equations and since the
differential equations are liner we obtain a general solution of
metric tensor by Green functions method and demonstrate that any
$f(X,Y,Z)$-theory corresponds to so-called \emph{Quadratic
Lagrangian}
($f(X,Y)\,=\,a_1R+a_2R^2+a_3R_{\alpha\beta}R^{\alpha\beta}$).
Initially the metric tensor is spherically symmetric and time
depending, but in this limit the dependence is missing (we need
the post-Newtonian order to fix a possible time dependence). So we
recover also a partial outcome about the Birkhoff theorem.

The metric potentials have two characteristic lengths depending on
the value of derivatives of $f$ with respect to curvature
invariants and only in GR are equal. The general solutions are
calculated for a point-like source and since the theory is linear,
the gravitational potential can be obtained for any matter
distribution.

With this general approach and by adding other curvature
invariants to action, this paper summarizes and generalizes the
topics of previously papers
\cite{PRD_mio,newtonian_limit_fR,newtonian_limit_R_Ric}.

\section{The Newtonian limit of $f(X,Y,Z)$-gravity}

Let us start with a general class of fourth order theories given
by the action

\begin{eqnarray}\label{HOGaction}\mathcal{A}=\int d^{4}x\sqrt{-g}\biggl[f(X,Y,Z)+\mathcal{X}\mathcal{L}_m\biggr]
\end{eqnarray}
where $f$ is an unspecified function of curvature invariants
$X$, $Y$ and $Z$.
The term $\mathcal{L}_m$ is the minimally coupled ordinary matter
contribution. In the metric approach, the field equations are
obtained by varying (\ref{HOGaction}) with respect to
$g_{\mu\nu}$. We get

\begin{eqnarray}\label{fieldequationHOG}
H_{\mu\nu}\,=\,&&f_XR_{\mu\nu}-\frac{f}{2}g_{\mu\nu}-f_{X;\mu\nu}+g_{\mu\nu}\Box
f_X+2f_Y{R_\mu}^\alpha
R_{\alpha\nu}-2[f_Y{R^\alpha}_{(\mu}]_{;\nu)\alpha}+\Box[f_YR_{\mu\nu}]+[f_YR_{\alpha\beta}]^{;\alpha\beta}g_{\mu\nu}+
\nonumber\\\nonumber\\&&+2f_ZR_{\mu\alpha
\beta\gamma}{R_{\nu}}^{\alpha\beta\gamma}-4[f_Z{{R_\mu}^{\alpha\beta}}_\nu]_{;\alpha\beta}\,=\,
\mathcal{X}\,T_{\mu\nu}
\end{eqnarray}
where
$T_{\mu\nu}\,=\,-\frac{1}{\sqrt{-g}}\frac{\delta(\sqrt{-g}\mathcal{L}_m)}{\delta
g^{\mu\nu}}$ is the the energy-momentum tensor of matter,
$f_X\,=\,\frac{df}{dX}$, $f_Y\,=\,\frac{df}{dY}$,
$f_Z\,=\,\frac{df}{dZ}$, $\Box={{}_{;\sigma}}^{;\sigma}$, and
$\mathcal{X}\,=\,8\pi G$\footnote{Here we use the convention
$c\,=\,1$.}. The conventions for Ricci's tensor is
$R_{\mu\nu}={R^\sigma}_{\mu\sigma\nu}$ while for the Riemann
tensor is
${R^\alpha}_{\beta\mu\nu}=\Gamma^\alpha_{\beta\nu,\mu}+...$. The
affinities are the usual Christoffel's symbols of the metric:
$\Gamma^\mu_{\alpha\beta}=\frac{1}{2}g^{\mu\sigma}(g_{\alpha\sigma,\beta}+g_{\beta\sigma,\alpha}
-g_{\alpha\beta,\sigma})$. The adopted signature is $(+---)$ (see
for the details \cite{landau}). The trace of field equations
(\ref{fieldequationHOG}) is the following

\begin{eqnarray}\label{tracefieldequationHOG}
H\,=\,g^{\alpha\beta}H_{\alpha\beta}\,=\,f_XR+2f_YR_{\alpha\beta}R^{\alpha\beta}+2f_ZR_{\alpha\beta\gamma\delta}
R^{\alpha\beta\gamma\delta}-2f+\Box[3
f_X+f_YR]+2[(f_Y+2f_Z)R^{\alpha\beta}]_{;\alpha\beta}\,=\,\mathcal{X}\,T
\end{eqnarray}
where $T\,=\,T^{\sigma}_{\,\,\,\,\,\sigma}$ is the trace of
energy-momentum tensor.

The paradigm of Newtonian limit is starting from a develop of
metric tensor (and of all additional quantities in the theory)
with respect to dimensionless quantity $v$ and considering only
first term of $tt$- and $ij$-component of metric tensor
$g_{\mu\nu}$ (for details see \cite{newtonian_limit_fR}). The
develop of metric tensor is as follows

\begin{eqnarray}\label{metric_tensor_PPN}
  g_{\mu\nu}\,\sim\,\begin{pmatrix}
  1+g^{(2)}_{tt}(t,\mathbf{x})+g^{(4)}_{tt}(t,\mathbf{x})+\dots & g^{(3)}_{ti}(t,\mathbf{x})+\dots \\
  g^{(3)}_{ti}(t,\mathbf{x})+\dots & -\delta_{ij}+g^{(2)}_{ij}(t,\mathbf{x})+\dots\end{pmatrix}
\end{eqnarray}

The set of coordinates\footnote{The Greek index runs between $0$
and $3$; the Latin index between $1$ and $3$.} adopted is
$x^\mu\,=\,(t,x^1,x^2,x^3)$. The curvature invariants $X$, $Y$,
$Z$ become

\begin{eqnarray}
\left\{\begin{array}{ll}
X\,\sim\,X^{(2)}+X^{(4)}+\dots\\\\
Y\,\sim\,Y^{(4)}+Y^{(6)}+\dots\\\\
Z\,\sim\,Z^{(4)}+Z^{(6)}\dots
\end{array}\right.
\end{eqnarray}
The function $f$ can be developed as

\begin{eqnarray}
f(X,Y,Z)\,&\sim&\,f(0)+f_X(0)X^{(2)}+\frac{1}{2}f_{XX}(0){X^{(2)}}^2+f_X(0)X^{(4)}+f_Y(0)Y^{(4)}+f_Z(0)Z^{(4)}+\dots
\end{eqnarray}
and analogous relations for partial derivatives of $f$ are
obtained. From lowest order of field equations
(\ref{fieldequationHOG}) we have

\begin{eqnarray}\label{PPN-field-equation-general-theory-fR-O0}
f(0)\,=\,0
\end{eqnarray}
Not only in $f(R)$-gravity \cite{newtonian_limit_fR,spher_symm_fR}
but also in $f(X,Y,Z)$-theory a missing cosmological component in
the action (\ref{HOGaction}) implies that the space-time is
asymptotically Minkowskian. The equations (\ref{fieldequationHOG})
and (\ref{tracefieldequationHOG}) at $\mathcal{O}(2)$ - order
become\footnote{We used the properties:
${R_{\alpha\beta}}^{;\alpha\beta}\,=\,\frac{1}{2}\Box R$ and
${{R_\mu}^{\alpha\beta}}_{\nu;\alpha\beta}\,=\,{{R_\mu}^\alpha}_{;\nu\alpha}-\Box
R_{\mu\nu}$.}

\begin{eqnarray}\label{NL-field-equation}
\left\{\begin{array}{ll}
H^{(2)}_{tt}\,=\,f_X(0)R^{(2)}_{tt}-[f_Y(0)+4f_Z(0)]\triangle
R^{(2)}_{tt}
-\frac{f_X(0)}{2}X^{(2)}-[f_{XX}(0)+\frac{f_Y(0)}{2}]\triangle
X^{(2)}\,=\,\mathcal{X}\,T^{(0)}_{tt}\\\\
H^{(2)}_{ij}\,=\,f_X(0)R^{(2)}_{ij}-[f_Y(0)+4f_Z(0)]\triangle
R^{(2)}_{ij}
+\frac{f_X(0)}{2}X^{(2)}\delta_{ij}+[f_{XX}(0)+\frac{f_Y(0)}{2}]\triangle
X^{(2)}\delta_{ij}-f_{XX}(0){X^{(2)}}_{,ij}+\\\\
\,\,\,\,\,\,\,\,\,\,\,\,\,\,\,\,\,\,\,\,\,\,+[f_Y(0)+4f_Z(0)]R^{(2)}_{mi,jm}+f_Y(0)R^{(2)}_{mj,im}\,=\,0\\\\
H^{(2)}\,=\,-f_X(0)X^{(2)}-[3f_{XX}(0)+2f_Y(0)+2f_Z(0)]\triangle
X^{(2)}\,=\,\mathcal{X}\,T^{(0)}
\end{array}\right.
\end{eqnarray}
where $\triangle$ is the Laplacian in the flat space. By
introducing the quantities

\begin{eqnarray}\label{mass_definition}
\left\{\begin{array}{ll}
{m_1}^2\,\doteq\,-\frac{f_X(0)}{3f_{XX}(0)+2f_Y(0)+2f_Z(0)}\\\\
{m_2}^2\,\doteq\,\frac{f_X(0)}{f_Y(0)+4f_Z(0)]}
\end{array}\right.
\end{eqnarray}
we get three differential equations for curvature invariant
$X^{(2)}$, $tt$- and $ij$-component of Ricci tensor
$R^{(2)}_{\mu\nu}$

\begin{eqnarray}\label{NL-field-equation_2}
\left\{\begin{array}{ll}
(\triangle-{m_2}^2)R^{(2)}_{tt}+\biggl[\frac{{m_2}^2}{2}-\frac{{m_1}^2+2{m_2}^2}{6{m_1}^2}\triangle\biggr]
X^{(2)}\,=\,-\frac{{m_2}^2\mathcal{X}}{f_X(0)}\,T^{(0)}_{tt}\\\\
(\triangle-{m_2}^2)R^{(2)}_{ij}+\biggl[\frac{{m_1}^2-{m_2}^2}{3{m_1}^2}\,
\partial^2_{ij}-\biggl(\frac{{m_2}^2}{2}-\frac{{m_1}^2+2{m_2}^2}{6{m_1}^2}\triangle\biggr)\delta_{ij}\biggr]
X^{(2)}\,=\,0\\\\
(\triangle-{m_1}^2)X^{(2)}\,=\,\frac{{m_1}^2\mathcal{X}}{f_X(0)}\,T^{(0)}
\end{array}\right.
\end{eqnarray}
We note that in the case of $f(R)$-theory we obtained a
characteristic length (${m_1}^{-1}$) on the which the Ricci scalar
evolves, but in $f(X,Y,Z)$-theory we have an additional
characteristic length (${m_2}^{-1}$) on the which the Ricci tensor
evolves. The solution for curvature invariant $X^{(2)}$ in third
line of (\ref{NL-field-equation_2}) is

\begin{eqnarray}\label{scalar_invariant_sol_gen}
X^{(2)}(t,\textbf{x})\,=\,\frac{{m_1}^2\mathcal{X}}{f_X(0)}\int
d^3\mathbf{x}'\mathcal{G}_1(\mathbf{x},\mathbf{x}')T^{(0)}(t,\mathbf{x}')
\end{eqnarray}
where $\mathcal{G}_1(\mathbf{x},\mathbf{x}')$ is the Green
function of field operator $\triangle-{m_1}^2$. The solution for
$g^{(2)}_{tt}$, by remembering
$R^{(2)}_{tt}\,=\,\frac{1}{2}\triangle g^{(2)}_{tt}$, is the
following

\begin{eqnarray}\label{tt_component_sol_gen}
g^{(2)}_{tt}(t,\textbf{x})\,=\,\frac{1}{2\pi}\int
d^3\mathbf{x}'d^3\mathbf{x}''\frac{\mathcal{G}_2(\mathbf{x}',\mathbf{x}'')}{|\mathbf{x}-\mathbf{x}'|}\biggl[
\frac{{m_2}^2\mathcal{X}}{f_X(0)}T^{(0)}_{tt}(t,\mathbf{x}'')-\frac{({m_1}^2+2{m_2}^2)\mathcal{X}}{6f_X(0)}\,T^{(0)}
t,\mathbf{x}'')+\frac{{m_2}^2-{m_1}^2}{6}X^{(2)}(t,\mathbf{x}'')\biggr]
\end{eqnarray}
where $\mathcal{G}_2(\mathbf{x},\mathbf{x}')$ is the Green
function of field operator $\triangle-{m_2}^2$. The expression
(\ref{tt_component_sol_gen}) is the "modified" gravitational
potential (here we have a factor 2) for $f(X,Y,Z)$-gravity. The
solution for the gravitational potential $g^{(2)}_{tt}/2$ has a
Yukawa-like behaviors (\cite{newtonian_limit_fR}) depending by a
characteristic lengths on whose it evolves.

The $ij$-component of Ricci tensor in terms of metric tensor
(\ref{metric_tensor_PPN}) is

\begin{eqnarray}\label{ricci_tensor}
R^{(2)}_{ij}\,=\,\frac{1}{2}g^{(2)}_{ij,mm}-\frac{1}{2}g^{(2)}_{im,mj}-\frac{1}{2}g^{(2)}_{jm,mi}-\frac{1}{2}g^{(2)}_
{tt,ij}+\frac{1}{2}g^{(2)}_{mm,ij}
\end{eqnarray}
and if we use the harmonic gauge condition
($g^{\alpha\beta}\Gamma^{\mu}_{\alpha\beta}\,=\,0$) the
(\ref{ricci_tensor}) becomes (\cite{newtonian_limit_fR})
$R^{(2)}_{ij}|_{HG}\,=\,\frac{1}{2}g^{(2)}_{ij,mm}\,=\,\frac{1}{2}\triangle
g^{(2)}_{ij}$. The general solution for $g^{(2)}_{ij}$ from
(\ref{NL-field-equation_2}), in the harmonic gauge, is

\begin{eqnarray}
g^{(2)}_{ij}|_{HG}\,=\,\frac{1}{2\pi}\int
d^3\mathbf{x}'d^3\mathbf{x}''\frac{\mathcal{G}_2(\mathbf{x}',\mathbf{x}'')}{|\mathbf{x}-\mathbf{x}'|}
\biggl[\frac{{m_1}^2-{m_2}^2}{3{m_1}^2}\,
\partial^2_{i''j''}-\biggl(\frac{{m_2}^2}{2}-\frac{{m_1}^2+2{m_2}^2}{6{m_1}^2}\triangle_{\mathbf{x}''}\biggr)\delta_{ij}\biggr]
X^{(2)}(\mathbf{x}'')
\end{eqnarray}
While if we hypothesize
$g^{(2)}_{ij}\,=\,2\,\psi\,\delta_{ij}$\footnote{We choose a
system of isotropic coordinates.} we have
$R^{(2)}_{ij}\,=\,\triangle\psi\,\delta_{ij}+(\psi-\phi)_{,ij}$
and the second field equation of (\ref{NL-field-equation_2})
becomes

\begin{eqnarray}\label{NL-field-equationij}
\left\{\begin{array}{ll} \triangle\psi\,=\,\int
d^3\mathbf{x}'\mathcal{G}_2(\mathbf{x},\mathbf{x}')\biggl(\frac{{m_2}^2}{2}-\frac{{m_1}^2+2{m_2}^2}{6{m_1}^2}
\triangle_{\mathbf{x}'}\biggr)X^{(2)}(\mathbf{x}')\\\\
(\phi-\psi)_{,ij}\,=\, \frac{{m_1}^2-{m_2}^2}{3{m_1}^2}\int
d^3\mathbf{x}'\mathcal{G}_2(\mathbf{x},\mathbf{x}')\,
X^{(2)}_{,i'j'}(\mathbf{x}')
\end{array}\right.
\end{eqnarray}
Then the general solution for $g^{(2)}_{ij}$ from
(\ref{NL-field-equation_2}), without gauge condition and by using
the first line of (\ref{NL-field-equationij}), is

\begin{eqnarray}\label{solpsi}
g^{(2)}_{ij}\,=\,2\,\psi\,\delta_{ij}\,=\,-\frac{\delta_{ij}}{2\pi}\int
d^3\mathbf{x}'d^3\mathbf{x}''\frac{\mathcal{G}_2(\mathbf{x}',\mathbf{x}'')}{|\mathbf{x}-\mathbf{x}'|}
\biggl(\frac{{m_2}^2}{2}-\frac{{m_1}^2+2{m_2}^2}{6{m_1}^2}
\triangle_{\mathbf{x}''}\biggr)X^{(2)}(\mathbf{x}'')
\end{eqnarray}
and the second line of (\ref{NL-field-equationij}) is only a
constraint condition for metric potentials. In fact from its trace
we have

\begin{eqnarray}\label{cond}
\triangle(\phi-\psi)\,=\,\frac{{m_1}^2-{m_2}^2}{3{m_1}^2}\int
d^3\mathbf{x}'\mathcal{G}_2(\mathbf{x},\mathbf{x}')\,
\triangle_{\mathbf{x}'}X^{(2)}(\mathbf{x}')
\end{eqnarray}
and we can affirm that only in GR the metric potentials $\phi$ and
$\psi$ are equals.

\section{Considerations about point-like solution}

\begin{figure}[htbp]
  \centering
  \includegraphics[scale=1]{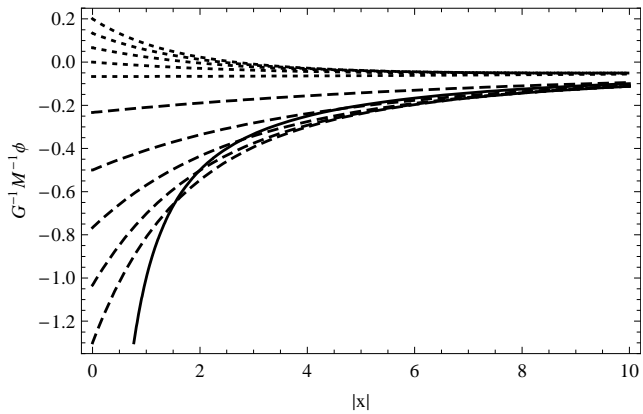}\\
  \caption{Plot of metric potential $\phi$ (\ref{sol_new_p}). $m_2\,=\,\xi\,m_1$ and $m_1\,=\,.1$ (dotted line),
  $m_1\,=\,\xi\,m_2$ and $m_2\,=\,.1$ (dashed line). The behavior of GR is shown by the solid line.
  The dimensionless quantity $\xi$ runs between $0\div 10$ with step $2$. The dimension of $m_1$ and $m_2$ is
  the inverse of length. We set $f_X(0)\,=\,1$.}
  \label{plotpotential}
\end{figure}

\begin{figure}[htbp]
  \centering
  \includegraphics[scale=1]{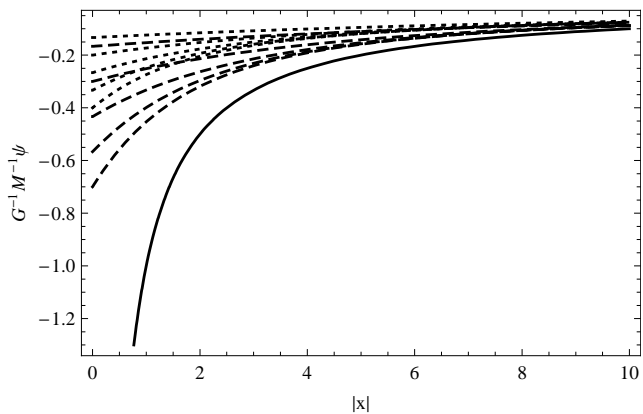}\\
  \caption{Plot of metric potential $\psi$ (\ref{sol_new_p_psi}). $m_2\,=\,\xi\,m_1$ and $m_1\,=\,.1$ (dotted line),
  $m_1\,=\,\xi\,m_2$ and $m_2\,=\,.1$ (dashed line). The behavior of GR is shown by the solid line.
  The dimensionless quantity $\xi$ runs between $0\div 10$ with step $2$. The dimension of $m_1$ and $m_2$ is
  the inverse of length. We set $f_X(0)\,=\,1$.}
  \label{plotpotential_2}
\end{figure}

Let us consider a point-like source with mass $M$. Then we have $
T^{(0)}_{tt}(t,\mathbf{x})\,=\,T^{(0)}(t,\mathbf{x})\,=\,M\delta(\mathbf{x})$,
and if we choose ${m_1}^2\,>\,0$ and ${m_2}^2\,>\,0$, the Green
functions $\mathcal{G}_i$ become
$\mathcal{G}_i(\mathbf{x},\mathbf{x}')\,=\,-\frac{1}{4\pi}\frac{e^{-m_i|\mathbf{x}-\mathbf{x}'|}}
{|\mathbf{x}-\mathbf{x}'|}$. The curvature invariant $X^{(2)}$
(\ref{scalar_invariant_sol_gen}) and the metric potentials $\phi$
(\ref{tt_component_sol_gen}) and $\psi$ (\ref{solpsi}) are

\begin{eqnarray}\label{sol_scalar}
X^{(2)}\,=\,
-\frac{r_g\,{m_1}^2}{f_X(0)}\frac{e^{-m_1|\mathbf{x}|}}{|\mathbf{x}|}
\end{eqnarray}

\begin{eqnarray}\label{sol_new_p}
\phi\,=\,-\frac{GM}{f_X(0)}\biggl[\frac{1}{|\textbf{x}|}
+\frac{1}{3}\frac{e^{-m_1|\mathbf{x}|}}{|\mathbf{x}|}
-\frac{4}{3}\frac{e^{-m_2|\mathbf{x}|}}{|\mathbf{x}|}\biggr]
\end{eqnarray}

\begin{eqnarray}\label{sol_new_p_psi}
\psi\,=\,-\frac{GM}{f_X(0)}\biggl[\frac{1}{|\textbf{x}|}
-\frac{1}{3}\frac{e^{-m_1|\mathbf{x}|}}{|\mathbf{x}|}
-\frac{2}{3}\frac{e^{-m_2|\mathbf{x}|}}{|\mathbf{x}|}\biggr]
\end{eqnarray}
where $r_g\,=\,2GM$ is the Schwarzschild radius. The modified
gravitational potential by $f(R)$-theory is further modified by
the presence of functions of $R_{\alpha\beta}R^{\alpha\beta}$ and
$R_{\alpha\beta\gamma\delta}R^{\alpha\beta\gamma\delta}$. The
curvature invariant $X^{(2)}$ (the Ricci scalar) presents a
massive propagation and when $f(X,Y,Z)\rightarrow f(R)$ we find
the mass definition $m^2\,=\,-f'(R=0)/3f''(R=0)$
(\cite{newtonian_limit_fR,newtonian_limit_R_Ric}) and propagation
mode with $m_2$ disappear. Obviously the expressions
(\ref{sol_new_p}) and (\ref{sol_new_p_psi}) satisfy the constraint
condition (\ref{cond}). In FIGs. \ref{plotpotential} and
\ref{plotpotential_2} we report the spatial behavior of metric
potentials for some values interval of parameters $m_1$ and $m_2$.


The same outcome can be obtained by considering the so-called
\emph{Quadratic Lagrangian}
$\mathcal{L}\,=\,\sqrt{-g}(a_1\,R+a_2\,R^2+a_3\,R_{\alpha\beta}R^{\alpha\beta})$
where $a_1$, $a_2$ and $a_3$ are constants. In this case
\cite{newtonian_limit_R_Ric} we found two characteristic lengths
$\biggl|\frac{2(3a_2+a_3)}{a_1}\biggr|^{1/2}$,
$\biggl|\frac{a_3}{a_1}\biggr|^{1/2}$ and the Newtonian limit of
theory implied as solution the equations (\ref{sol_new_p}) and
(\ref{sol_new_p_psi}). We can affirm, then, the Newtonian limit of
any $f(X,Y,Z)$-theory can be reinterpreted by introducing the
\emph{Quadratic Lagrangian} and the coefficients have to satisfy
the following relations

\begin{eqnarray}\label{equivalence}
a_1\,=\,f_X(0),\,\,\,\,\,\,\,\,\,a_2\,=\,\frac{1}{2}f_{XX}(0)-f_Z(0),\,\,\,\,\,\,\,\,\,
a_3\,=\,f_Y(0)+4f_Z(0)
\end{eqnarray}
A first considerations about (\ref{equivalence}) is regarding the
characteristic lengths induced by $f(X,Y,Z)$-theory. The second
length ${m_2}^{-1}$ is originated from the presence, in the
Lagrangian, of Ricci and Riemann tensor square, but also a theory
containing only Ricci tensor square could show the same outcome
(it is successful replacing the coefficients $a_i$ of
\emph{Quadratic Lagrangian} or renaming the function $f$).
Obviously the same is valid also with the Riemann  tensor square
alone. Then a such modification of theory enables a massive
propagation of Ricci Tensor and, as it is well known in the
literature, a substitution of Ricci Scalar with any function of
Ricci scalar enables a massive propagation of Ricci scalar. We
can, then, affirm that an hypothesis of Lagrangian containing any
function of only Ricci scalar and Ricci tensor square is not
restrictive and only the experimental constraints can fix the
arbitrary parameters.

A second consideration is starting from the Gauss - Bonnet
invariant defined by the relation $G_{GB}\,=\,X^2-4Y+Z$
\cite{dewitt_book}. In fact the induced field equations satisfy in
four dimensions the following condition

\begin{eqnarray}\label{fieldequationGB}
H^{GB}_{\mu\nu}\,=\,H^{X^2}_{\mu\nu}-4H^{Y}_{\mu\nu}+H^{Z}_{\mu\nu}\,=\,0
\end{eqnarray}
and by substituting them at Newtonian level
($H^Z_{tt}\,\sim\,-4\triangle R^{(2)}_{tt}$) in the equations
(\ref{fieldequationHOG}) we find the field equations (ever at
Newtonian Level) of \emph{Quadratic Lagrangian}.

A third and last consideration is about the solutions
(\ref{sol_new_p}) and (\ref{sol_new_p_psi}). When we perform the
limit in the origin $|\mathbf{x}|\,=\,0$ we don't have the
divergency. In fact we find

\begin{eqnarray}\label{divergence}
\lim_{|\mathbf{x}|\rightarrow 0
}\phi\,=\,\frac{m_1-4m_2}{3},\,\,\,\,\,\,\,\,\,\,\,\,\lim_{|\mathbf{x}|\rightarrow
0 }\psi\,=\,-\frac{m_1+2m_2}{3}
\end{eqnarray}
and only if we remove in the action (\ref{HOGaction}) the
dependence on the Ricci square or Riemann square we get the known
divergence of GR. For a physical interpretation of solution
(\ref{sol_new_p}) we must impose the condition $m_1-4m_2\,<\,0$ to
have a potential well with a negative minimum in
$|\mathbf{x}|\,=\,0$ and $m_1\,<\,m_2$ to have a negative profile
of potential (see FIG. \ref{plotpotential}). Then, if we suppose
$f_X(0)\,>\,0$, we get a constraint on the derivatives of $f$ with
respect to curvature invariants

\begin{eqnarray}\label{condition}
f_{XX}(0)+f_Y(0)+2f_Z(0)\,<\,0
\end{eqnarray}
In the case of $f(R)$-gravity ($f_Y(0)\,=\,f_Z(0)\,=\,0$) we
reobtain the same condition among the first and second derivatives
of $f$ \cite{PRD_mio,newtonian_limit_fR,newtonian_limit_R_Ric}.

\section{Conclusions}\label{conclusions}

In this paper the theory of Gravity induced by a most general
fourth order theory obtained by using all curvature invariants has
been considered. By adding these curvature invariants, this paper
summarized and generalized the topics of previously papers
\cite{PRD_mio,newtonian_limit_fR,newtonian_limit_R_Ric}. In fact
for a generic $f(X,Y,Z)$-theory, at Newtonian level, it is
successful considering only the so-called \emph{Quadratic
Lagrangian}. All contributions to field equation due by curvature
invariant Riemann square can be expressed by other two curvature
invariants (Ricci tensor square and Ricci scalar square) via
Gauss-Bonnet invariant.

The spherically symmetric solutions of metric tensor at
$\mathcal{O}$(2)-order show a Yukawa-like dependence only by two
characteristic lengths and not by three (because we have three
curvature invariants in the action). No gauge condition has been
considered and the solution of $ij$-component of metric tensor is
general. Is general also the solution of $tt$-component of metric
tensor, since it is gauge free (but only at Newtonian order).

Furthermore, generally it has been shown that for a
$f(X,Y,Z)$-gravity, but the same is valid also for $f(R)$-gravity,
the metric potentials are not equal. Only in the limit
$f\,\rightarrow\,R$ we obtain the outcome of GR. This aspect with
the consequences of Birkhoff and Gauss theorem are the principal
differences between a fourth order gravity and GR.

The general solutions are calculated for a point-like source and
since the theory is linear, the gravitational potential can be
obtained for any matter distribution. The metric potentials don't
have the divergency in $|\mathbf{x}|\,=\,0$ for a point-like
source, and by requiring a right physical interpretation of
solution we get a constraint on the derivatives of $f$ with
respect to curvature invariants. The constraint condition is
compatible with respect to one obtained for $f(R)$-gravity.
Besides a such class of theory have free parameters and only the
experimental evidence can fix them.

\section{Acknowledges}

The author would like to thank prof. Salvatore Capozziello for his
useful discussions and suggestions.

\end{document}